\documentstyle[epsfig,12pt]{article}

\newcommand{\bce}{\begin{center}} 
\newcommand{\ece}{\end{center}}
\newcommand{\beq}{\begin{equation}}
\newcommand{\eeq}{\end{equation}}
\newcommand{\bea}{\vspace{0.25cm}\begin{eqnarray}}
\newcommand{\eea}{\end{eqnarray}}

\newcommand{\brho}{\mbox{\boldmath $\rho$}}

\newcommand{\br}{{\bf r}}

\newcommand{\ba}{\begin{array}}
\newcommand{\ea}{\end{array}}

\newcommand{\ket}[1]{| {#1} \rangle}
\newcommand{\bra}[1]{\langle {#1} |}

\newcommand{\doublespace}{
    \renewcommand{\baselinestretch}{1.6}\large\normalsize}

\newcommand{\bb}{{\bf b}}
\def\lsim{\mathrel{\rlap{\lower4pt\hbox{\hskip1pt$\sim$}}
    \raise1pt\hbox{$<$}}}         %less than or approx. symbol
\def\gsim{\mathrel{\rlap{\lower4pt\hbox{\hskip1pt$\sim$}}
    \raise1pt\hbox{$>$}}}         %greater than or approx. symbol

\def\Pom{{\bf I\!P}}

\def\beq{\begin{equation}}
\def\endeq{\end{equation}}
\def\arr{\begin{eqnarray}}
\def\endarr{\end{eqnarray}}
\makeindex

  \advance\oddsidemargin  by -1in
  \advance\evensidemargin by -1in
  \advance\topmargin      by -0.5in
\begin{document}

\vspace{2.0cm}

\begin{flushright}

\end{flushright}

\vspace{1.0cm}

\begin{center}
{\Large \bf 
Unitarity constraints for DIS off nuclei:\\ predictions for electron-ion colliders.
}

\vspace{1.0cm}

{\large\bf N.N.~Nikolaev$^{a,b}$, W.~Sch\"afer$^{c}$
B.G.~Zakharov$^{b}$ and V.R.~Zoller$^{d}$}

\vspace{1.0cm}
{\sl
$^{a}$IKP(Theorie), FZ J{\"u}lich, J{\"u}lich, Germany\medskip\\
$^{b}$L.D. Landau Institute for Theoretical Physics, Moscow 117940, Russia
\medskip\\
$^{c}$Institute of Nuclear Physics PAN, PL-31-342 Cracow, Poland\medskip\\
$^{d}$ Institute for Theoretical and Experimental Physics, Moscow 117218, Russia.}
\vspace{1.0cm}\\
{ \bf Abstract }\\
\end{center}
Future electron-ion 
colliders (eIC) will focus on the unitarity properties
of deep inelastic scattering (DIS) in the limit of strong nuclear
absorption. Strong nuclear shadowing and a
large abundance of coherent diffraction are the most striking 
consequences of unitarity, and here we report quantitative 
predictions for these effects in the kinematical range of the
planned eIC.

\doublespace
\pagebreak

%---------------------------------------------------

%-------------------------------------------

{\bf {1. Introduction.}}\\

In  deep-inelastic scattering off nuclei at the Bjorken variable
\beq
x\ll x_A= {1\over {m_N R_A}} =0.15 A^{-1/3},
\label{eq:XA}
\eeq
where $R_A$ is the radius of the target nucleus of mass number A 
and $m_N$ is
the nucleus mass, unitarity driven nuclear shadowing (NS)
 becomes important \cite{NZfusion,Saturation,Nonlinear}. 
It comes along with the large-rapidity-gap coherent diffractive 
DIS (LRG DIS) when the target nucleus remains in the ground state.
For strongly absorptive nuclei the unitarity condition was shown
to imply a paradoxically large, $\sim 50\%$, fraction of LRG 
DIS \cite{NZZdiffr}
(for the experimental confirmation of LRG DIS off nuclei
see \cite{Adams}, similar calculations 
were reported in the recent Ref. \cite{Kugeratski}).
It is precisely the s-channel unitarity condition which controls the 
interplay of the virtual and real pQCD radiative corrections
in the small-$x$ evolution of different nuclear observables
\cite{VirtualReal}.  
In this communication we report  quantitative predictions
for the small-$x$ evolution of NS and LRG DIS for the energy 
range of future eIC \cite{eRHIC}
which will test the unitarity properties of  hard scattering 
processes under strong nuclear absorption (saturation) . \\

{\bf {2. First iteration of the LL$(1/x)$ evolution for 
nuclear cross sections.}}\\  

We base our work on the color dipole (CD) approach to the Leading 
Log$(1/x)$ (LL$(1/x)$), 
or BFKL \cite{BFKL}, evolution of DIS \cite{NZZBFKL,NZ94}.
For nuclear targets a complete resummation of LL$(1/x)$ 
effects is as yet lacking, the Double
Leading Log \cite{BGZ98}, large-$N_c$ Color Glass
Condensate \cite{Saturation} and the fan diagram  resummation
\cite{Kaidalov} approximations were considered in the literature
( for the review see 
\cite{Armesto},  the dominance of fan diagrams 
was questioned, though \cite{AGK}). Our point is that at energies
of the  planned
eIC \cite{eRHIC}, the $x$-dependence of  NS and LRG DIS
is practically exhausted by the 
first CD LL$(1/x)$ iteration which is calculable exactly
without invoking the large-$N_c$ 
approximation \cite{Saturation,Nonlinear}. Indeed,  the 
first CD LL$(1/x)$ iteration dominates in 
the $x$ region
\beq
\xi = \log {x_0 \over x } \lsim {1\over \Delta_{eff}}, 
\label{eq:REGION} 
\eeq
where $\Delta_{eff}\simeq 0.1-0.2$ is the exponent of the
local $x$-dependence of the proton structure function,
$F_{2p} \propto x^{-\Delta_{eff}}$ \cite{HERAdeltaPom}.

\begin{figure}[t]
\psfig{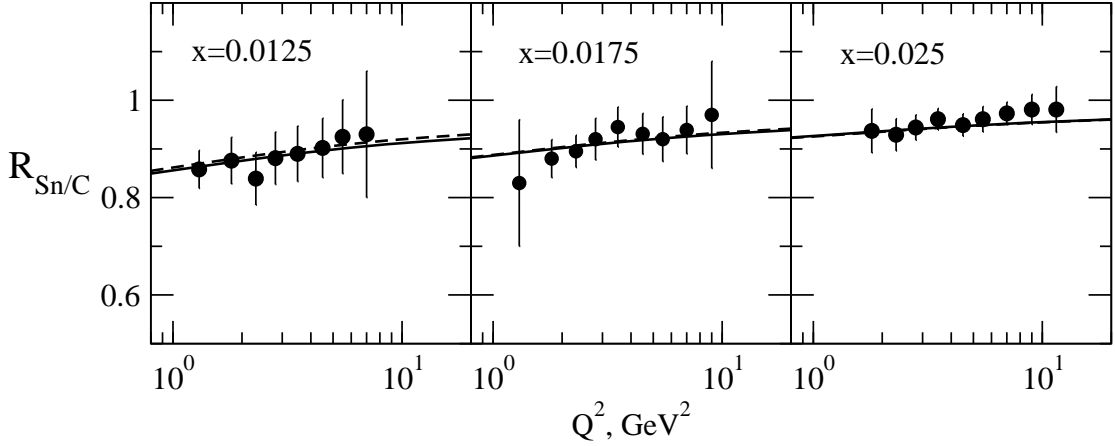}
%\vspace{1.5cm}\\
\caption{\small The predictions form CD LL$(1/x)$ evolution vs. the 
NMC data on the shadowing ratio $R_{Sn/C}$ \cite{NMCSn}.
The
solid lines represent the full shadowing from the $q\bar{q}$ and $q\bar{q}g$
Fock states, while the dashed lines show a shadowing at the 
$q\bar{q}$ level, see also a caption to Fig. \ref{fig:RNUC}. } 
\label{fig:NMC}
\end{figure}

When viewed in the laboratory frame, NS derives from the coherent 
interaction of $q\bar q, q\bar q g,...$
states. To the required accuracy, the Fock state expansion 
of the physical photon $|\gamma^*\rangle$ reads
$%\bea
|\gamma^*\rangle=\sqrt{Z_g} \Psi_{q\bar q}|q\bar q\rangle+
\Phi_{q\bar q g}|q\bar q g\rangle,
$%\label{eq:FOCK}
%\eea
where $\Psi_{q\bar q}$ and 
$\Phi_{q\bar q g}$ 
are the light-cone wave functions (WF's)
of the $q\bar q$ and $q\bar q g$ states,
$\sqrt{Z_g}$ is the renormalization of the  
$q\bar q$ state by the virtual radiative corrections 
for the $q\bar q g$ state. For soft gluons
the 3-parton WF takes the 
factorized form 
$\Phi_{q\bar q g}=\Psi_{q\bar q}\{\Psi_{qg}-\Psi_{{\bar q}g}\}$ \cite{NZZBFKL,NZ94}. 
The nuclear coherency condition reads \cite{NZfusion}
\beq
%{M^2 + Q^2 \over 2m_N\nu} = 
{x/\beta} \lsim x_A,
\label{eq:COHERENCY}
\eeq
where  $x=Q^2/2m_N\nu$ is the Bjorken variable, 
$\nu$ is the energy of the photon, 
$M$ is the invariant mass of the multiparton Fock state
and $\beta= Q^2/(M^2 + Q^2)$. 

In the CD
BFKL-Regge phenomenology of DIS one usually formulates the boundary 
condition 
at $x= x_0=0.03$ \cite{BFKLRegge}. 
For extremely heavy nuclei $x_A\ll x_0$ 
and the LL$(1/x)$ evolution starts at $x=x_A$ 
with the boundary condition formulated in terms of the free-nucleon
quantities  LL$(1/x)$-evolved from $x_0$ down to $x_A$. 
At small $x$ satisfying the condition (\ref{eq:COHERENCY}) the 
color dipoles $\{\br_n\}$ in the multiparton state 
are conserved in the interaction process.  At the boundary $x=x_A$ and for
$A\gg 1$, the nuclear 
$\textsf{S}$-matrices equal \cite{Glauber,Gribov} 
\beq 
\textsf{S}_{n}(x_A,\bb,\{\br\}_n) = \exp[-{1\over 2} \sigma_n(x_A,\{\br\}_n)T(\bb)],
\label{eq:SMATRIXC}
\eeq 
where  $\sigma_n(x_A,\{\br\}_n)$ is the free-nucleon CD cross section for the
$n$-parton state, and \cite{NZ91}
\beq
\sigma_{2,A}(x_A,\br )=2\int d^2\bb[1 - \textsf{S}_{2}(x_A,\bb,\br)].
\label{eq:SIGMA0}
\eeq
Here 
$T(\bb)=\int_{-\infty}^{+\infty} dz n_A(\sqrt{z^2+\bb^2}),
$
is the optical  thickness of a nucleus  
at an impact parameter $\bb$, the nuclear matter density $n_A(\br)$
is normalized as  
$\int d^3\br n(\br)=A $.

\begin{figure}[t]
\psfig{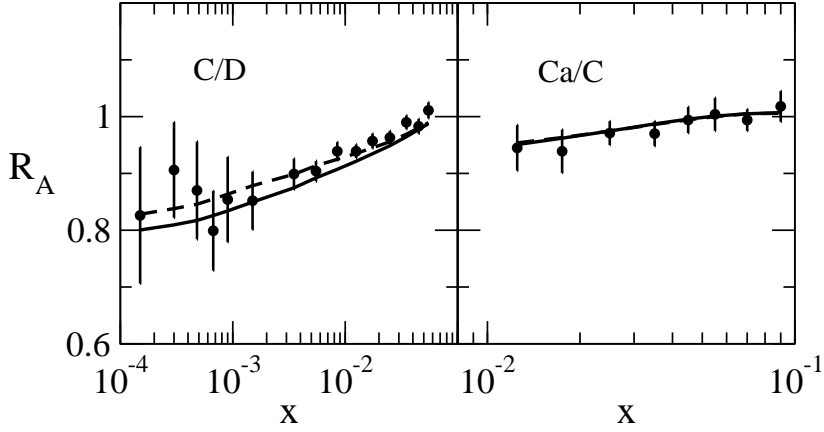}
\caption{\small The same as in Fig. 1 for the 
NMC data on the shadowing ratio $R_{C/D}$ \cite{NMCCarb}
and $R_{Ca/C}$  \cite{NMCCalc} 
as a function of two correlated  variables $x$ and $Q^2$ 
(for details of $x-Q^2$ correspondence see \cite{NMCCarb},\cite{NMCCalc}). 
 } 
\label{fig:NMC2}
\end{figure}

A quite common extension of Eq.(\ref{eq:SIGMA0}) to  $x\ll x_A$ with the
BFKL-evolved  $\sigma_2(x,\br)$ is completely unwarranted. Instead one 
must evolve the nuclear $\textsf{S}$-matrix. Specifically,
after the gluon variables have been properly integrated out, the effect of the
extra gluon in the Fock state expansion for the incident photon boils down
to the renormalization of the $\textsf{S}$-matrix and nuclear cross section 
for the $q\bar{q}$ CD \cite{NZZBFKL,NZ94,VirtualReal},:
\bea
{\partial{\textsf S}_2(x,\br;\bb) \over \partial \xi }
&=&\int d^2\brho  |\psi(\brho) -\psi(\brho+\br)|^2  \Big[
{\textsf S}_{3}(x,\brho,\br;\bb) - {\textsf S}_{2}(x,\br;\bb) \Big], \nonumber\\
\sigma_A(x,\br) &=& \sigma_{2,A}(x_A,\br ) +\sigma^{(1)}_A(x,\br ), \nonumber\\
\sigma^{(1)}_A(x,\br)& =& 2\log\left({x_A\over x}\right)
\int d^2\bb\int d^2\brho  |\psi(\brho) -\psi(\brho+\br)|^2\nonumber\\
&\times& \Big[ {\textsf S}_{2}(x_A,\br;\bb)- {\textsf S}_{3}(x_A,\brho,\br;\bb)\Big],\nonumber\\
\sigma_3(x,\br,\brho) &=& {N_c^2 \over N_c^2-1}\Big[\sigma_2(x,\brho) +
\sigma_2(x,\brho + \br)\Big] - {1 \over N_c^2-1} \sigma_2(x,\br)\label{eq:DGDXI}
\eea
where $\rho$ is the $qg$ dipole and 
\beq
\psi(\brho)={\sqrt{C_F\alpha_S}\over \pi}\cdot {\brho\over \rho^2 }\cdot {\rho
  \over R_c} K_1({\rho\over R_c})
\label{eq:PSI}
\eeq
is the radial WF of the $qg$ state with the Debye screening
of infrared gluons \cite{NZZBFKL,NZ94}.
The virtual photoabsorption cross section is 
an expectation value $\sigma_A(x,Q^2) =\langle q\bar q|\sigma_A(x,\br)|q\bar
q\rangle\,.$  
 
\begin{figure}[t]
\psfig{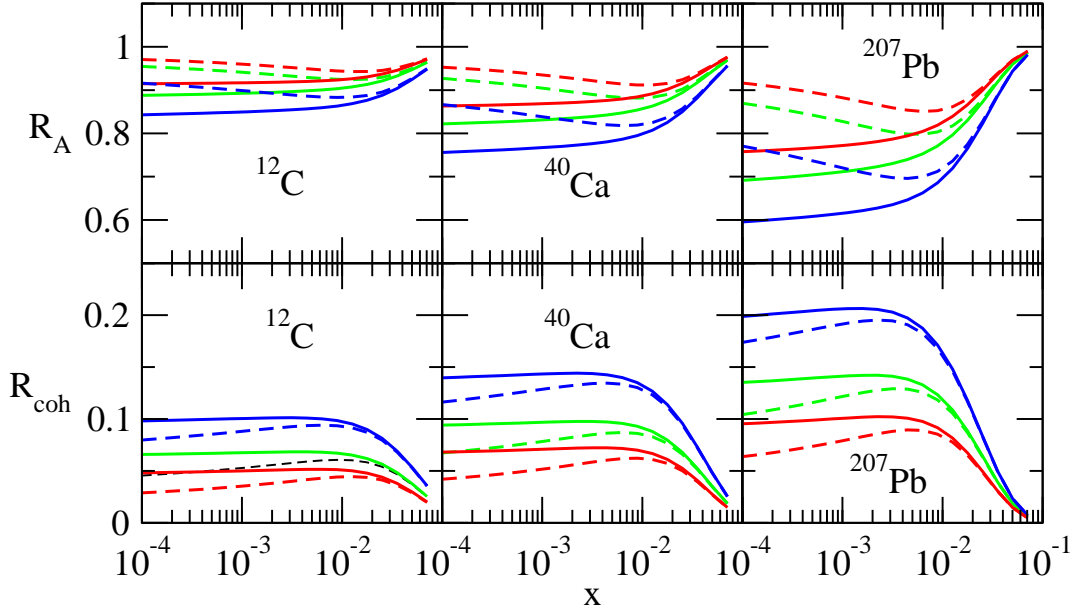}
\caption{\small  The upper panels: predictions from CD LL$(1/x)$ evolution
for NS ratio $R_A$ for C, Ca and Pb nuclei as a function of $x$ at  
$Q^2=1,5,20$ GeV$^2$, the bottom to top curves, respectively.
The solid lines show the combined NS (\ref{eq:Rshad}) from the $q\bar{q}$ and 
$q\bar{q}g$ Fock states
of the photon, while the dashed lines 
represent NS at the $q\bar{q}$ level, i.e., the term
  $R_{2,A}=1-[A\sigma_{2,N}(x_A,Q^2)-\sigma_{2,A}(x_A,Q^2)]/A\sigma_N(x)$ 
in the expansion  (\ref{eq:Rshad}). The lower panels: the same as above for
the fraction of DIS which is LRG coherent diffraction, $R_{coh}$.
 Dashed lines represent the contribution from
 the low-mass  $q\bar q$ excitations, $\sigma_A^{LM}$, the sum of 
the low-mass and $3\Pom$ high-mass terms is shown by solid lines.} 
\label{fig:RNUC}
\end{figure}

The shadowing ratio $R_A(x,Q^2)$, decomposed into the $q\bar{q}$ and $q\bar{q}g$ 
contributions, equals  
\bea
R_A(x,Q^2)={\sigma_A(x,Q^2)\over  A\sigma_N(x,Q^2)} = 1 &-&
{ A\sigma_{2,N}(x_A,Q^2)-\sigma_{2,A}(x_A,Q^2) \over
  A\sigma_N(x,Q^2)}\nonumber\\
& -& 
\log\left({x_A\over x}\right) {A\sigma^{(1)}_{N}(x_A,Q^2)-\sigma^{(1)}_{A}(x_A,Q^2) \over  A\sigma_N(x,Q^2)}, 
\label{eq:Rshad}
\eea 
where  $\sigma_N(x,Q^2)=\sigma_{2,N}(x_A,Q^2)+
\log(x_A/x)\sigma^{(1)}_{N}(x_A,Q^2)$ is 
the LL$(1/x)$-evolved free-nucleon cross section. 
Our interest is in the well 
evolved shadowing at $x\ll x_A$, the onset of NS at $x \gsim x_A$ must be treated
within the light-cone Green function technique \cite{BGZ98,BGZ96}, here
we show only the gross features of the large-$x$ suppression of NS
following the prescriptions from Refs. \cite{NZ91,Karmanov} and the 
$\beta$-dependence of LRG DIS as predicted in \cite{NZ92} and
confirmed in the HERA experiments \cite{ZEUSdiff}. We use the free-nucleon
CD 
cross section tested against the
experimental data from HERA, it is described in the Appendix.
The nuclear density 
parameters are taken from the compilation 
\cite{DEVRIES}.
In Fig. \ref{fig:NMC} we compare  the predictions from the first 
LL$(1/x)$ iteration with the NMC data \cite{NMCSn} on
the ratio of $R_{Sn/C}(x,Q^2)={R_{Sn}(x,Q^2)/R_{C}(x,Q^2)}$.
The results for $R_{C/D}$ \cite{NMCCarb} and $R_{Ca/C}$ \cite{NMCCalc}
 are shown in Fig. \ref{fig:NMC2}.
The agreement with the experimental data is good, but in a limited 
region of $x,Q^2$ the contribution from
the shadowing of the $q\bar{q}g$ Fock states is still  small. 
A much broader range of $x$ and $Q^2$ can be covered at eIC
\cite{eRHIC}, and in Fig. \ref{fig:RNUC}
we show our predictions for $R_A(x,Q^2)$ at eIC. 

{\bf{3. LL$(1/x)$ evolution for LRG DIS}}\\

The LL$(1/x)$ evolution  of fully inclusive forward 
LRG DIS off a free nucleon, starting from $x_\Pom=x_0$,
has the Fock state expansion \cite{NZ92,NZ94}
\bea
\left.{{d\sigma_N^D}(x,x_0)\over dt}\right|_{t=0}={1\over 16\pi}
\left[\langle q\bar{q}|\sigma_2^2(x_0)|q\bar{q}\rangle
+\langle q\bar{q}g|\sigma_3^2(x_0)-\sigma_2^2(x_0)|q\bar{q}g\rangle+...\right] \, ,
\label{eq:NucleonDD}
\eea
where $t$ is the $(p,p')$ momentum transfer squared. 
Upon the integration over the gluon momentum
the $q\bar{q}g$ term  
gives rise to the LL$(1/x)$
evolution of LRG DIS. Its 
decomposition \cite{qqff}
\beq
\sigma_3^2(x_0)-\sigma_2^2(x_0) = [\sigma_3(x_0)-\sigma_2(x_0)]^2 +
2\sigma_2(x_0) (\sigma_3(x_0) -\sigma_2(x_0))
\label{eq:DECOMP}
\eeq
splits the LL$(1/x)$ evolution into the real production of
high-mass (triple-pomeron, $3\Pom$) states governed by
$[\sigma_3(x_0)-\sigma_2(x_0)]^2$ 
and the virtual
pQCD radiative correction to the Born excitation of low-mass, $M^2\sim Q^2$,
states
described by the first term of eq.(\ref{eq:DECOMP}).

For heavy nuclei we consider coherent diffractive DIS  per 
unit area in the impact parameter space, $d\sigma_A^D(x,x_0)/d^2\bb$
with the boundary $x_0=x_A$. At fixed $x\ll x_A$, the small-mass 
$q\bar{q}$ diffraction enters at $x_\Pom\approx x$,
while the small-$x$ evolution of the high-mass diffraction 
starts with the rapidity gap variable $x_\Pom = x_A$. To 
the LL$(1/x)$ approximation, integrating out the gluon
variables, one can cast it in the color dipole
form \cite{qqff} $d\sigma_A^D(x,x_A,\bb)/d^2\bb =\bra{q\bar{q}} 
\Sigma_A^D(x,x_A,\bb,\br)
\ket{q\bar{q}}$,
where, with certain reservations \cite{POMglue}, one can dub
$\Sigma_A^D(x,x_A,\bb,\br)$ a CD cross section for the pomeron target.
Extending the
decomposition (\ref{eq:DECOMP}) to nuclear targets, 
for the   
high-mass excitations, i.e.,  $3\Pom$ at $x\ll x_A$, one finds \cite{qqff}
\bea
{d\sigma_A^{3\Pom}(x,x_A,\bb,\br)\over d^2\bb}&=&
\langle q\bar{q}g|\left[\textsf{S}_{2}(x_A,\bb,\br)-
\textsf{S}_{3}(x_A,\bb,\brho,\br)\right]^2|q\bar{q}g\rangle = 
\bra{q\bar{q}} \Sigma_A^{3\Pom}(x,x_A,\bb,\br)
\ket{q\bar{q}},\nonumber\\
\Sigma_A^{3\Pom}(x,x_A,\bb,\br)&=& \log\left({x_A\over x}\right)
\int d^2\brho  |\psi(\brho) 
-\psi(\brho+\br)|^2
\left[\textsf{S}_{2}(x_A,\bb,\br)-
\textsf{S}_{3}(x_A,\bb,\brho,\br)\right]^2.
\label{eq:Nucleus3P}
\eea
Analogously,  for the  low-mass (LM) 
excitations ${d\sigma_A^{LM}/d^2\bb}=
\bra{q\bar{q}} \Sigma_A^{LM}\ket{q\bar{q}}$ and
\bea
\Sigma_A^{LM}(x,x_A,\bb,\br)&=& \left[1-\textsf{S}_{2}(x_A,\bb,\br)\right]^2
+2\log\left({x_A\over x}\right)\left[1-\textsf{S}_{2}(x_A,\bb,\br)\right]\nonumber\\
&\times&\int d^2\brho  |\psi(\brho) 
-\psi(\brho+\br)|^2
\left[\textsf{S}_{2}(x_A,\bb,\br)-
\textsf{S}_{3}(x_A,\bb,\brho,\br)\right].
\label{eq:NucleusQQ}
\eea

\begin{figure}[t]
\psfig{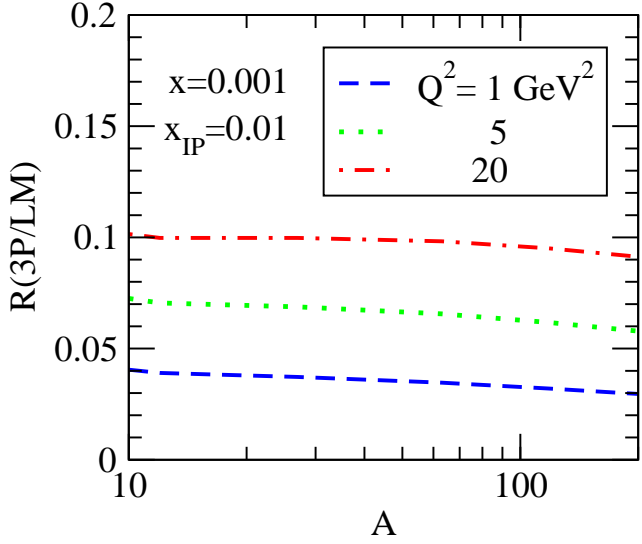}
%\vspace{1.5cm}
\caption{\small The predicted nuclear mass-number dependence 
of the ratio, $R(3\Pom/LM)$, of the high-mass to low-mass diffractive DIS at
several values of $Q^2$.  }
\label{fig:3PLM}
\end{figure}

 A fraction of DIS which is coherent diffraction,
$
 R_{coh}(x,Q^2)={\sigma_A^D(x,x_A,Q^2)/\sigma_A(x,x_A,Q^2)} $,
is shown in 
Fig.\ref{fig:RNUC}.
 The large-$x$ suppression of $
R_{coh}(x,Q^2)$
by the nuclear form
factor
is evaluated with the mass spectrum \cite{NZ92,ZEUSdiff}.
For the realistic dipole cross section even the lead nucleus is still
a grey one: the predicted $R_{coh}$ is substantially smaller than
the black disc result  $R_{coh}=0.5$ \cite{NZZdiffr}.

The total cross section of the low-mass diffraction  $\sigma_A^{LM}(x,Q^2)$,
i.e.,  DIS off the large-$\beta$ valence $q\bar{q}$ state of the pomeron,
is saturated by the contribution from $M^2 \sim Q^2$. 
The strength of $3\Pom$ 
diffraction, or DIS off the small-$\beta$ 
sea in the pomeron,  is conveniently measured by 
\beq
G_{3\Pom}(x_{\Pom})= {\partial \over \partial \xi}
\int d^2\bb  \bra{q\bar{q}} \Sigma_A^{3\Pom}(x,x_A,\bb,\br)
\ket{q\bar{q}} = (M^2+Q^2){d\sigma_A^{3\Pom}(x_{\Pom})\over dM^2}. 
\label{eq:A3P}
\eeq
In Fig. \ref{fig:3PLM} we show the ratio 
$
R(3\Pom/LM)={G_{3\Pom}(x_\Pom)/\sigma_A^{LM}(x)},
$
where we take $x=10^{-3}$ and $x_\Pom = x/\beta=10^{-2}$, i.e., $\beta=0.1$,
where the diffractive excitation of the $q\bar{q}g$ states takes
over the excitation of the  $q\bar{q}$ states (\cite{NZ92}, for
the experimental data see \cite{ZEUSdiff}). At such a value of 
$x_\Pom$ the nuclear form factor effects can be neglected. 
The rise of $R(3\Pom/LM)$ with $Q^2$
is a standard pQCD scaling violation growth of the sea parton density.
Subtleties of the scaling violations at small $Q^2$ close to
the nuclear saturation scale $Q^2_A$ will be discussed elsewhere.
 
The suppression of $R(3\Pom/LM)$ for heavy nuclei, 
evident in Fig. \ref{fig:3PLM}, is a unitarity
effect. Indeed, the nuclear absorption in the integrand 
of triple-pomeron cross section, $\Sigma_A^{3\Pom}(x,x_A,\bb,\br)$,
is stronger than that of $\Sigma_A^{LM}(x,x_A,\bb,\br)$, and it
is steadily enhanced for large $A$. 
\\

{\bf{5. Summary.}}\\

Based on the color dipole approach which correctly 
reproduces the experimental
data on the proton structure function measured at HERA and
the NMC data on nuclear shadowing,
we reported the quantitative predictions for the  LL$(1/x)$ 
evolution of the nuclear shadowing  and coherent diffraction 
dissociation off nuclei in the kinematical range to be covered
at future electron-ion colliders. We explicitly separated the 
contributions to both observables from the low-mass and 
the triple-pomeron high-mass diffractive states,
including the virtual radiative correction to the low-mass diffraction.
 Regarding the saturation properties, even the 
lead nucleus is still a grey one: the fraction of DIS which is diffractive
is substantially below the black disc result  $R_{coh}=0.5$. 
The gross features of the predicted $x$ and $Q^2$ dependence of
nuclear shadowing do not change much from the carbon to lead target;
we also predict a suppression of the high-mass vs. low-mass 
coherent diffractive DIS on heavy nuclei.

{\bf Acknowledgments.}
This work has been partly supported by 
the DFG grant 436 RUS 17/82/06. VRZ acknowledges also   partial 
 support from the RFBR grant 06-02-16905-a.\\

{\bf{Appendix. Boundary condition for the LL$(1/x)$ evolution.}}\\

\begin{figure}[t]
\psfig{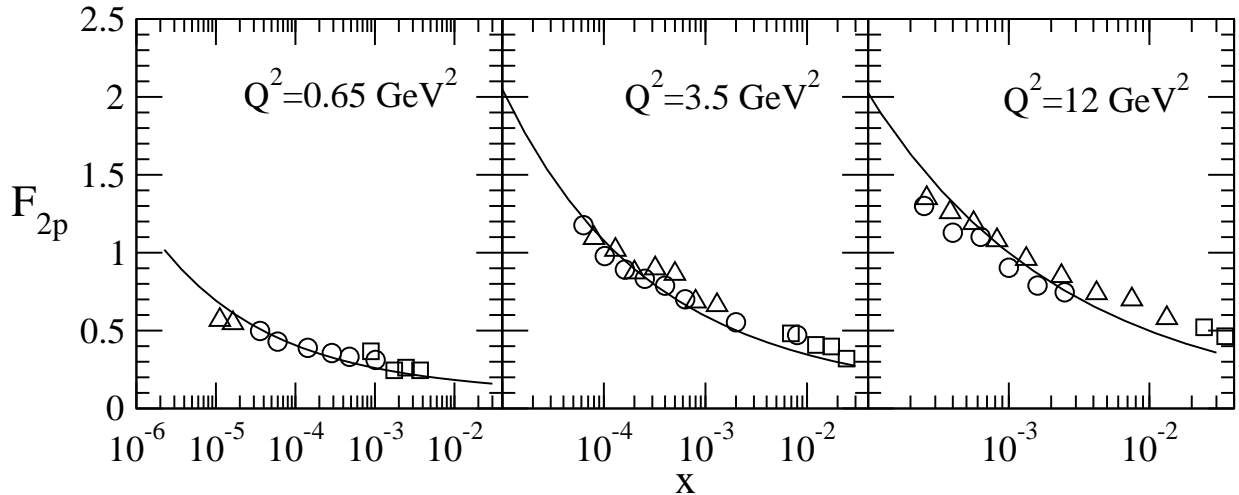}
\caption{\small 
The proton structure function $F_{2p}(x,Q^2)$  predicted by 
the BFKL-evolved color dipole cross section (\ref{eq:SIG003}) vs.
the experimental data from ZEUS (open 
circles, \cite{ZEUSSF}), H1 (triangles,\cite{H1SF})  
and E665 (squares, \cite{E665SF}) Collaborations.}
\label{fig:f2p}
\end{figure} 

The realistic input $\sigma(x_0,\br)$ must be defined for all scales $r$,
from small to non-perturbative large.  Motivated by our 
successful BFKL-Regge color-dipole phenomenology of
DIS \cite{BFKLRegge,qqff}, at $x_0=0.03$ and in the limited range of
$r\gsim 10^{-3}$ fm of the practical interest, we take 
\beq
\sigma(x_0,r)=\sum_{i=1}^3{\sigma_ia_i(r)
\over {1+a_i(r)}},
\label{eq:SIG003}
\eeq
where $a_i=(r/r_0)^{2+\gamma_i}$, $r_0=1$ fm, $\sigma_1=22.5$ mb, 
$\sigma_2=20$ mb, $\sigma_3=8$ mb, $\gamma_1=0.15$,$\gamma_2=0.8$, 
$\gamma_3=2$. Its LL$(1/x)$ evolution is  described by the Color Dipole 
BFKL equation \cite{NZZBFKL,NZ94} with the infrared freezing of the
one-loop, 3-flavor QCD coupling,
\beq
\alpha_S(q^2)={4\pi\over {9\log\left[{{(q^2+q^2_f)}/\Lambda^2_{QCD}}\right]}},
\label{eq:ALPHAS}
\eeq
where $q_f\approx 0.7$ GeV and  $\Lambda_{QCD}=0.3$
GeV. A new observation is that at the expense of somewhat smaller 
Debye screening radius, $R_c =0.2$ fm 
vs. $R_c =0.275$ fm in \cite{NZZBFKL,NZ94}, the LL$(1/x)$ evolution
of the boundary condition (\ref{eq:SIG003}) gives a good description of
the experimental data from HERA (see Fig \ref{fig:f2p}) without splitting the boundary condition
into the BFKL-evolving perturbative, and non-evolving, non-perturbative, 
components.

\end{document}